\documentclass[a4paper]{article}
\usepackage{graphicx,amsmath,amssymb,amsthm,mathtools,parskip,url,tikz,hyperref}
\usepackage[left=2cm, right=2cm, top=1.5cm, bottom=2.5cm]{geometry}
\usepackage{setspace}

\title{Strassen's algorithm via orbit flip graphs}
\author{Christian Ikenmeyer and Jakob Moosbauer
\\
{\small
University of Warwick}\\
\small{
\texttt{$\{$christian.ikenmeyer, jakob.moosbauer$\}$@warwick.ac.uk}
}}
\date{March 2025}

\newcommand{\ot}{\otimes}
\newcommand{\tensor}{{\textstyle\bigotimes}}
\newcommand{\IF}{\mathbb{F}}
\newcommand{\IN}{\mathbb{N}}

\begin{document}
\raggedbottom

\maketitle

\begin{abstract}
We give a short proof for Strassen's result that the rank of the 2 by 2 matrix multiplication tensor is at most~7. The proof requires no calculations and also no pattern matching or other type of nontrivial verification, and is based solely on properties of a specific order 6 group action. Our proof is based on the recent combination of flip graph algorithms and symmetries.
\end{abstract}

Let $\IF$ be any field, and let $\IF^{2\times 2}$ have the standard basis $\{e_{0,0},e_{0,1},e_{1,0},e_{1,1}\}$.
The $2 \times 2$ matrix multiplication tensor is defined as
$M_2 = \sum_{i,j,k=0}^1 e_{i,j} \ot e_{j,k} \ot e_{k,i} \in \tensor^3(\IF^{2\times 2})$, see \cite[(14.20)]{BCS97}.
The \emph{rank} $R(t)$
for $t \in \tensor^3(\IF^{2\times 2})$
is defined as the smallest number $r\in \IN$ such that there exist $u_{i},v_{i},w_{i}\in \IF^{2\times 2}, i=1,\ldots, r$, with
$t = \sum_{i=1}^{r} u_{i} \ot v_{i} \ot w_{i}$.
Strassen \cite{Str69} proved that $R(M_2)\leq 7$, and we reprove this result.

\begin{spacing}{1.075}
The cyclic group $C_3 = \langle\pi \mid \pi^3=1\rangle$ acts linearly on $\tensor^3(\IF^{2\times 2})$ via $\pi(u\ot v \ot w) = w \ot u \ot v$
and $\IF$-linear extension.
For $i \in \{0,1\}$, let 
$i' := 1-i$.
The cyclic group $C_2= \langle\tau \mid \tau^2=1\rangle$ acts linearly on $\tensor^3(\IF^{2\times 2})$ via
$\tau(e_{i_1,i_2}\ot e_{j_1,j_2} \ot e_{k_1,k_2})
= e_{i'_1,i'_2}\ot e_{j'_1,j'_2} \ot e_{k'_1,k'_2}$
and $\IF$-linear extension.
Both actions commute, so that we have an action of $G := C_3\times C_2$.
We write $G\cdot t := \sum_{g\in G} g(t)$, which is a notation introduced in \cite{BILR19}.
For a standard basis vector $t$ it is easy to visualize
$G\cdot t = G\cdot \tau(t)$ and $G\cdot t = G\cdot \pi(t)$: For $\tau$ we
connect pairs of inverted indices $i$ and $i'$ with red lines, and for $\pi$ we
illustrate the cyclic position rotation with blue lines:
\end{spacing}
\begin{tikzpicture}
\node at (-5,0) {
\begin{minipage}{5cm}
\vspace{-0.3cm}
\begin{align*}
&&G\cdot e_{0,1}\ot e_{1,1} \ot e_{0,0}
\\[0.2cm]
&&=G\cdot e_{1,0}\ot e_{0,0} \ot e_{1,1}
\end{align*}
\end{minipage}
};
\node at (0,-0.1) {and};
\node at (5,0) {
\begin{minipage}{5cm}
\vspace{-0.3cm}
\begin{align*}
&&G\cdot e_{0,1}\ot e_{1,1} \ot e_{0,0}
\\[0.2cm]
&&=G\cdot e_{0,0}\ot e_{0,1} \ot e_{1,1}
\end{align*}
\end{minipage}
};
\draw[thick,red] (-5.27,-0.36) -- (-5.27,0.04);
\draw[thick,red] (-5.04,-0.36) -- (-5.04,0.04);
\draw[thick,red] (-4.29,-0.36) -- (-4.29,0.04);
\draw[thick,red] (-4.06,-0.36) -- (-4.06,0.04);
\draw[thick,red] (-3.32,-0.36) -- (-3.32,0.04);
\draw[thick,red] (-3.09,-0.36) -- (-3.09,0.04);
\draw[thick,blue!50!white] (5.8,-0.30) -- (4.83,0);
\draw[thick,blue!50!white] (4.83,-0.30) -- (6.77,0);
\draw[thick,blue!50!white] (6.77,-0.30) -- (5.8,0);
\end{tikzpicture}

The proof of $R(M_2)\leq 7$ now goes as follows.
There are two different kinds of summands in the 8 summands of~$M_2$: 6 of them form $G\cdot e_{1,0} \ot e_{0,1} \ot e_{1,1}$. For the remaining 2 summands, we can get their sum from a rank~1 tensor with 8 terms and subtracting the 6 superfluous terms as follows:
\[
e_{0,0}\ot e_{0,0} \ot e_{0,0} + e_{1,1}\ot e_{1,1} \ot e_{1,1} \ = \ (e_{0,0}+e_{1,1})\ot (e_{0,0}+e_{1,1})\ot (e_{0,0}+e_{1,1})
- G \cdot e_{0,0}\ot e_{0,0} \ot e_{1,1}.
\]
So far, by naively looking at the summand structure, we have obtained a rank $1+6+6=13$ decomposition for~$M_2$.
We now observe:

\begin{tikzpicture}
\node at (-4.7,1.45) {$M_2 \, - \, (e_{0,0}+e_{1,1})\ot (e_{0,0}+e_{1,1})\ot (e_{0,0}+e_{1,1})$};
\node at (0,0) {\begin{minipage}{\textwidth}
\begin{alignat*}{6}
&=
  G \cdot e_{1,0} \ot e_{0,1} \ot e_{1,1}
 && && 
 &&- G \cdot e_{0,0} \ot e_{0,0} \ot e_{1,1}
\\
&=
 G \cdot e_{1,0} \ot e_{0,1} \ot e_{1,1}
&{} + \big(&- G \cdot e_{0,1} \ot e_{1,1} \ot e_{0,0}&&+\hspace{4.5pt} G \cdot e_{0,1} \ot e_{1,1} \ot e_{0,0} \big)
&&- G \cdot e_{0,0} \ot e_{0,0} \ot e_{1,1}
\\[0.2cm]
&=\mathrlap{
 \big(G \cdot e_{1,0} \ot e_{0,1} \ot e_{1,1}}
&&- G \cdot e_{1,0} \ot e_{0,0} \ot e_{1,1}\big)
&& + \big(G \cdot e_{0,0} \ot e_{0,1} \ot e_{1,1} 
&&- G \cdot e_{0,0} \ot e_{0,0} \ot e_{1,1}\big)
\\
&=
 \mathrlap{\hspace{1.5cm} G \cdot e_{1,0} \ot (e_{0,1}-e_{0,0}) \ot e_{1,1}}
&&
&&+ \mathrlap{\hspace{1.5cm}G \cdot e_{0,0} \ot (e_{0,1}-e_{0,0}) \ot e_{1,1}}
&&
\\
&=
 \mathrlap{\hspace{4.5cm}G \cdot (e_{1,0}+e_{0,0}) \ot (e_{0,1}-e_{0,0}) \ot e_{1,1},}
\end{alignat*}
\end{minipage}};
\draw[thick,red] (-2.27,-0.26) -- (-2.27,0.14);
\draw[thick,red] (-2.04,-0.26) -- (-2.04,0.14);
\draw[thick,red] (-1.29,-0.26) -- (-1.29,0.14);
\draw[thick,red] (-1.06,-0.26) -- (-1.06,0.14);
\draw[thick,red] (-0.32,-0.26) -- (-0.32,0.14);
\draw[thick,red] (-0.09,-0.26) -- (-0.09,0.14);
\draw[thick,blue!50!white] (2.6,-0.20) -- (1.63,0.1);
\draw[thick,blue!50!white] (1.63,-0.20) -- (3.57,0.1);
\draw[thick,blue!50!white] (3.57,-0.20) -- (2.6,0.1);
\end{tikzpicture}
which is of rank at most 6.
Hence, $M_2$ has rank at most~7.

\section*{Context and related work}
For $n\in\IN$, let $M_n := \sum_{i,j,k=0}^{n-1} e_{i,j} \ot e_{j,k} \ot e_{k,i}$.
The matrix multiplication exponent $\omega = \inf\{\log_{n}(R(M_n))\}$ governs the running time of the algorithmic task of multiplying two matrices and of other tasks in linear algebra, for example solving systems of linear equations, see \cite[\S15.1, \S16]{BCS97}.
In order to understand $R(M_n)$ for large~$n$, significant research effort has focused on understanding $R(M_n)$ for small values of $n$, with several papers focusing on $n=2$ \cite{Bre70, Gas71, Fid72, BD73, Pat74, Laf75, Mak75, Yuv78, BC85, Cla88, Cha86, Ale97, GK00, Pat09, GM17, IL19, CILO19}, since no satisfying short proof of $R(M_2)\leq 7$ has been found.
The verification of these proofs usually requires simple but somewhat lengthy calculations, and careful attention to the details has to be paid.
The weaker analogous result for the tensor border rank
also gives $\omega\leq\log_2(7)$ and
can be obtained from Lickteig's analysis of defective secant varieties, see \cite{Lic85}, \cite[\S3.8, \S9]{Lan08}, which is significantly more involved than our proof.

Our new proof of $R(M_2)\leq 7$ is so short that it can hopefully be used
successfully as a blueprint for proofs for other small~$n$.  In fact, it does
arise from a machinery for proofs for small $n$, the flip graph algorithm
\cite{KM23}.  $R(M_n)$ for small values of $n$ has been studied for a long time
\cite{HK71,Lad76,JM86,Mak87,DIS11,Smi13,Sed17,HKS19,HKS21,SS21,Smi21,Fawetal22,Smi23,AIH24,KM25,Sed25}.
The recent flip graph algorithm works as follows: one starts with a simple but
expensive decomposition of $M_n$ and reduces its rank via a series of
simplifications.  Adding ideas from \cite{BILR19} (tensor rank decompositions of
$M_3$), this technique has recently been improved by considering the symmetry
groups $C_3$ and $C_3\times C_2$ to get new upper bounds on $R(M_5)$ and
$R(M_6)$, see \cite{MP25}.  Our proof can be interpreted as a refined version of
the output of this newest flip graph algorithm.  Note that \cite{BILR19} also
consider a group of order 6, but the group there is $C_3 \rtimes C_2 \simeq
S_3$.  The flip graph proof technique requires to specify a trivial but expensive solution,
a group (usually a subgroup of the symmetry group of $M_n$), and a sequence of
trivial simplifications.  As a consequence, our proof generates 7
summands instead of starting with them and verifying their correctness.
\cite{deG78} showed that any rank 7 decomposition is equivalent to Strassen's original decomposition.
It is
an open question which starting solutions and which groups can be used to get
satisfying short proofs for $M_n$, $n \geq 3$.

\bibliographystyle{alpha}
{\footnotesize
\bibliography{lit}
}
\end{document}